\documentclass[preprint,aps,showpacs,prb,superscriptaddress]{revtex4-1}
\usepackage{graphicx,color}
\usepackage{dcolumn}
\usepackage{bm}
\usepackage{amsmath}

\usepackage[T1]{fontenc}       % Use modern font encodings
\usepackage[english]{babel}
\usepackage[utf8x]{inputenc}
\usepackage{graphicx}
\usepackage{amsmath}
\usepackage{amssymb}
\usepackage{xspace}
\usepackage{hyperref}
\usepackage{natbib}

\begin{document}

\title{Effects of exchange interactions  on magnetic anisotropy and spin-dynamics  of  adatoms on metallic surfaces
}
\author{P. Ruiz-D\'{i}az}

\affiliation{Max-Planck-Institut f\"{u}r Mikrostrukturphysik Weinberg 2, 
D-06120 Halle, Germany}
\author{O. V. Stepanyuk}
\affiliation{Max-Planck-Institut f\"{u}r Mikrostrukturphysik Weinberg 2, 
D-06120 Halle, Germany}
\affiliation{Institution of Russian Academy of Sciences Dorodnicyn Computing Centre of RAS,
Vavilov st. 40, 119333 Moscow, Russia}
\author{V.~S.~Stepanyuk}
\affiliation{Max-Planck-Institut f\"{u}r Mikrostrukturphysik Weinberg 2, 
D-06120 Halle, Germany}

\date{\today}

\let\AAt\AA
\renewcommand{\AA}{\ifmmode{\textup{\AAt}}\else{\AAt}\fi}
\newcommand{\var}[1]{\ensuremath{#1}\xspace}
\newcommand{\unt}[1]{\ensuremath{\mathrm{#1}}\xspace}
\newcommand{\chem}[1]{\ensuremath{\rm #1}\xspace}
\newcommand{\mB}{{\rm \mu_B}}
\newcommand{\Exc}{\ensuremath{E_\mathrm{ex}}\xspace}
\newcommand{\Ema}{\ensuremath{E_\mathrm{MA}}\xspace}

\newcommand{\bra}[1]{\left\langle #1 \right\vert}
\newcommand{\ket}[1]{\left\vert #1 \right\rangle}
\newcommand{\bracket}[2]{\left\vert #1 \middle| #2 \right\rangle}
\newcommand{\matel}[3]{\left\langle #1 \middle\vert #2 \middle\vert #3 \right\rangle}
\newcommand{\abs}[1]{\left|#1\right|}

\begin{abstract}
 	Both quantum and classical behavior of single atomic spins on surfaces is determined by the local anisotropy of adatoms and their coupling to the immediate electronic
 	environment. Yet adatoms seldom reside on surfaces alone and it is generally acknowledged that substrated-mediated interactions can couple single spins among each other 
 	impacting their magnetic behavior. Here we show that also magnetic anisotropy, which is usually considered to be a constant determined by the local crystal field, can be 
 	extremely sensitive to such interactions. By the example of Co dimers on Cu(001) and Pt(001) surfaces we highlight the intricate interplay of exchange coupling and magnetic
 	anisotropy providing a much sought possibility to tune the latter through deliberate adjustment of the adatoms' separation. As a technologically relevant implication we
 	demonstrate the impact of such emergent non-local anisotropy on the hysterectic properties of single-atom magnetization curves.
\end{abstract}

\maketitle

%%%%%%%%%%%%%%%%%%%%%%%%%%%%%%%%%%%%%%%%%%%%%%%%%%%%%%%%%%%%%%%%%%%%%
%% Main part

%\section{Introduction}

	For a while now the interest of surface science community has been keenly focused on nano-magnets \cite{Gatteschi2006} being of significance from both technological \cite{Leuenberger2001,Troiani2005,Khajetoorians2011} and basic magnetism \cite{Brune2013,Balashov2009,Miyamachi2013} standpoints. With scanning probe microscopy having reached unprecedented ubiquity as a tool of choice for state-of-the-art basic surface science studies, the field of view of most studies has narrowed down to single adatoms or single-molecule-magnets (SMM). The key parameters defining both quantum and classical behavior of such systems, i.e. magnetic moment, anisotropy energy (\Ema) and coupling to environment, can be determined for individual spins based on inelastic tunneling spectroscopy (IETS) \cite{Hirjibehedin2007,Otte2008,Donati2013,Rau2014} and single-atom magnetometry \cite{Zhou2010,Wiebe2011,Khajetoorians2012}.

	To extract the above parameters by magnetization or IETS measurements and to model the systems one usually resorts to simplified semi-empiric Hamiltonians \cite{Hirjibehedin2007,Otte2008,Zhou2010,Wiebe2011,Khajetoorians2012,Rau2014}, which in most cases do a fairly good job of describing the experimental observation. To account for dynamic effects (e.g. Kondo and quantum spin tunneling) or anisotropic interactions more elaborate formulations \cite{Kikoin2012,Zitko2009,Ferron2015,Yan2015} have been resorted to. Yet to date, all of them have one thing in common -- they treat magnetic anisotropy as an intrinsic constant parameter of the adatom or SMM, usually defined by the local adsorption geometry, the resulting ligand field and the spin-orbital coupling in the system\cite{Otte2008,Otte2009,Khajetoorians2012,Rau2014}.

	At the same time, in real life adatoms seldom reside on the surface alone. For example, it is well known that interatomic exchange interaction with neighboring impurities, for one, can strongly affect the magnetic order in the system across distances of several nanometers via direct \cite{Khajetoorians2012} and substrate-mediated Ruderman-Kittel-Kasuya-Yosida (RKKY) \cite{Wahl2007,Brovko2008prl,Wiebe2011} or super-exchange \cite{Lin2012a,Abdurakhmanova2013,Umbach2012} type interactions and can be further directionally anisotropic \cite{Zhou2010}. Recently, on the continuous quest for finding ways to deliberately tune magnetic properties of surface spins \cite{Ruiz-Diaz2013,Ruiz-Diaz2014a,Brovko2014}, it has been found that magnetic anisotropy can be modified by changing the coupling of the adsorbate spin to the substrate electron bath \cite{Oberg2014,Jacobson2015nc}, scanning probe-tip interaction \cite{Yan2014} and hydrogenation \cite{Jacobson2015nc,Dubout2015}, but also by the 
introduction of neighboring adatoms in the immediate vicinity \cite{Bryant2013} via the changes in the surface relaxation induced thereby.

	Careful analysis of experimental data, however, reveals scattered hints that also intermediate and long-range interaction of adatoms can have an impact on the magnetic anisotropies in the system. For example, the variation of anisotropies of Co adatoms on semi-insulating CuN islands on Cu(001) surface \cite{Oberg2014} can only partially be explained by the variation of the coupling to the substrate across the island if the presence of other Co adatoms on the same island is neglected. Also, in general, the scatter of anisotropy values tends to be higher in experiments with higher density of adsorbates \cite{Oberg2014,Dubout2015} and on substrate exhibiting stronger RKKY properties (metallic substrates versus semi-insulating ones) \cite{Gambardella2003,Wiebe2011,Dubout2015}. More isolated impurities on decoupling substrates \cite{Otte2008,Rau2014,Donati2013,Jacobson2015nc}, on the contrary, usually exhibit much smaller error-bars of anisotropy values.

	In the present letter we address this issue and show that electronic interaction between impurities does not only manifest itself in exchange coupling variation across distances of several nanometers but can equally strongly affect the anisotropy of individual adatoms. Using the prototypical system of Co adatoms on (001) surfaces of Cu and Pt and a combination of first-principles and Kinetic-Monte-Carlo (KMC) approaches we demonstrate how adjusting the distance between individual spins on the surface one can tune their local anisotropies and thus their response to external stimuli, the latter being illustrated by interatomic-distance dependent single-atom magnetization curves.

%\section{Numerical approach}

\begin{figure}
		\center{\includegraphics{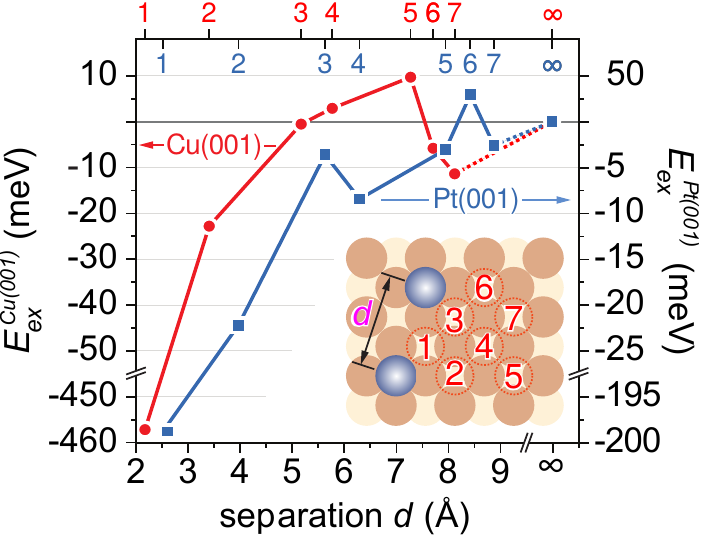}} \caption{Exchange interaction between Co adatoms deposited on Cu(001) (red circles) and Pt(001) (blue squares) surfaces as a function of adatom separation $d$. The rightmost point denoted $\infty$ corresponds to a non-interacting dimer or single adatom. Numbers on the top axis indicate the nearest-neighbor index corresponding to the numbers in the inset (here the color-code is: dark blue - Co, medium brown - first surface layer of Cu/Pt, light brown - second layer  of Cu/Pt).} \label{fig:01:exch}
	\end{figure}

	Our calculations are based on the projector augmented wave (PAW) method as implemented in the Vienna \textit{Ab-initio} Simulation Package (VASP) code\cite{Kresse1993,Kresse1996}. To exchange interaction is treated within the generalized gradient approximation in the Perdew-Becke-Ernzerhof form \cite{Perdew1996}. The substrate is modeled by a large ($10 \times 8$ atoms) cubic super-cell slab having five-layers thickness and $16~\AA$ of vacuum space. The considered Co dimer configurations are shown in the inset of Figure~\ref{fig:01:exch}. A $3 \times 4$ $k$-mesh is employed to sample the Brillouin zone with a plane-wave energy grid cut-off at $450$ eV. First, scalar-relativistic calculations are carried out for the structural optimization and then fully-relativistic ones to determine the \Ema. Spin-orbit coupling \cite{Blonski2009b} and dipolar corrections \cite{Neugebauer1992} are taken into account in the latter calculations. Geometrical relaxations are
	performed until the forces acting on the atoms 
        become less than $10^{-2}~\unt{eV/\AA}$ and the change in the total energy between two successive electronic steps is smaller than $10^{-7}$~eV. The \Ema of the dimers is calculated from the difference in the sum of the one-electron Kohn-Sham band energies at a fixed potential (magnetic force theorem) \cite{Daalderop1994,Lehnert2010,Blonski2009}. To simulate the magnetization curves Kinetic-Monte-Carlo method is employed as described in Refs.~\citenum{Smirnov2009} and \citenum{Li2006}.

%\section{Exchange interaction}

	We begin by revisiting the well-known phenomenon of RKKY-mediated exchange interaction of magnetic adatoms (in our case Co) on metallic surfaces. Figure~\ref{fig:01:exch} depicts the exchange interaction energy $\Exc = E_\mathrm{FM} - E_\mathrm{AFM}$ (defined as the energy difference between ferro- (FM) and antiferro-magnetically (AFM) coupled states of the dimer) as a function of the separation distance $d$ between the Co adatoms on Cu(001) (red circles) and Pt(001) (blue rectangles) surfaces. In the case of Cu(001) surface, the results of our calculation are, as expected, in good semi-quantitative agreement with the results obtained in earlier non-relaxed calculations based on the Korringa-Kohn-Rostoker (KKR) Green function method \cite{Wahl2007}. Indeed, structural relaxations are significant up to second-nearest neighbors (NN) ($d\sim3.7~\AA$) only. For compact dimers the Co-Co bond length is significantly contracted with respect to the surface inter-site distance. The Co-Pt distance is also smaller 
than Co and Pt interlayer distance (about $17\%$ relaxation for Cu and $27\%$ for Pt). However, the in-plane relaxations are negligibly small for larger interatomic separations and even the relaxations towards the surface do not show any interplay with the observed magnetic order (which also explains the surprising match of our results and KKR calculations).
	
	A predominant FM coupling is obtained for most separations except a region of $5 < d < 7~\AA$ where a FM-to-AFM transition is observed \cite{Wahl2007}. All in all, the exchange interaction dependence on the separation distance $d$ shows a non-monotonous behavior in line with the RKKY physics \cite{Wahl2007}. For Co atoms on Pt(001) surface we obtain a similarly non-monotonous exchange interaction dependence on $d$ (blue squares in Figure~\ref{fig:01:exch}), which is comparable in strength to that on Cu(001) at larger separations and is weaker than on Cu(001) for compacter dimers due to the larger Pt lattice constant.

%\section{Separation dependent anisotropy in dimers}

	\begin{figure}
		\centering\includegraphics{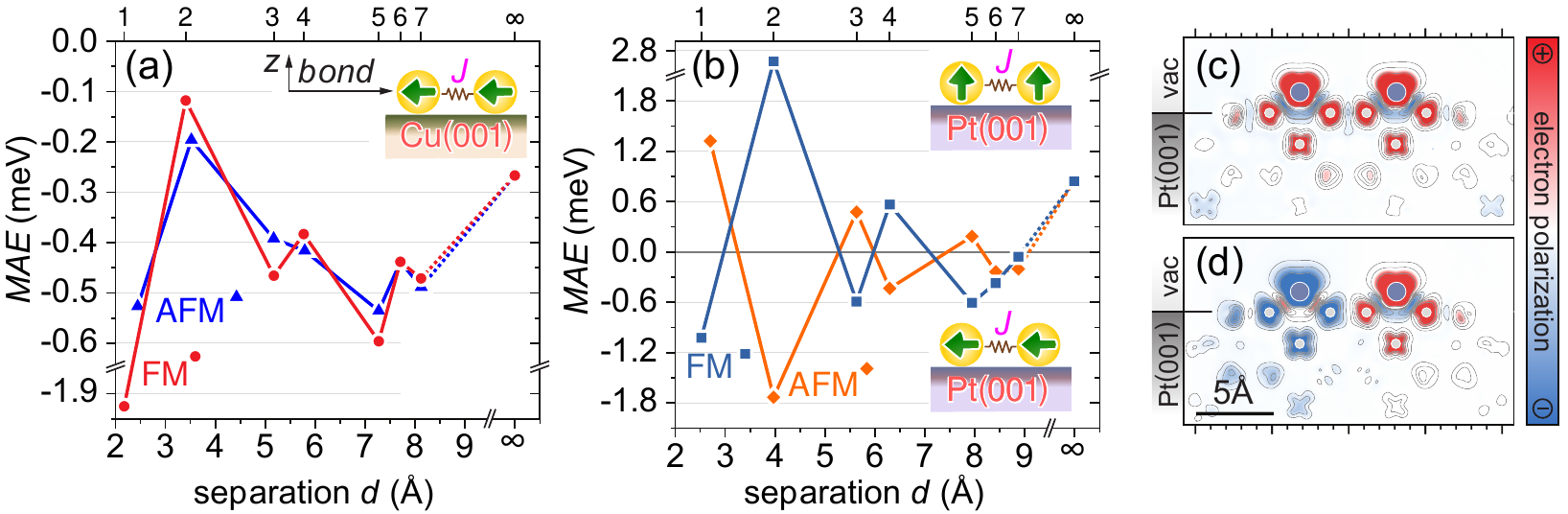}
		\caption{Dependence of local per-atom magnetic anisotropy of (a) a Co dimer on a Cu(001) surface considering both ferromagnetic (FM, red circles) and antiferromagnetic (AFM, blue triangles) ordering of spins and the same for (b) a Co-dimer on a Pt(001) (FM in blue squares and AFM in orange rhombs). Positive \Ema values stand for out-of plane \Ema. The rightmost point denoted $\infty$ corresponds to a non-interacting dimer or single adatom. The spin polarization of electrons induced in the Pt(001) surface by the presence of a ferromagnetic (c) and antiferromagnetic (d) Co dimers at $6.28~\AA$ separation.} 
		\label{fig:02:maepol}
	\end{figure}

	Conventionally, knowing the $d$-dependent exchange coupling strengths presented above along with on-site anisotropy values for a single Co adatom on Cu or Pt surface would be 
	enough to fully describe the magnetic behavior of the system. However, if we take a look at the values of \Ema self-consistently calculated for FM coupled Co dimers on Cu(001) 
	surface presented in Figure~\ref{fig:02:maepol}(a) (red circles) we shall see that the values are far from being constant but rather exhibit a strong dependence on the 
	interatomic separation $d$. \Ema for a Co dimer $\Ema = E_\mathrm{z} - E_\mathrm{bond}$ (defined as the energy difference between the dimers with spins aligned along the normal
	to the surface and the bond axis) is non-monotonously changing in the range of $0.1-0.6~\unt{meV}$ with increasing $d$. The only exception is the compact dimer with stronger
	$\Ema = 1.94~\unt{meV}$).

	Doubtlessly, \Ema of the Co adatoms is related with local environment of the adsorbates. However, it proved rather difficult to put a handle on the electronic origin of \Ema. 
	The local density of states (LDOS) of Co adatoms exhibits slightly variations with regard to the magnetic order of the dimers. Only the orbital-resolved LDOS analysis is
	meticulous enough to provide  insights into both the magnetic order and distance-dependence of \Ema. A more detailed discussion of the subject can be found in the supplement.
	Here we shall only mention that using the second-order perturbation formula \cite{Wang1993b}, the \Ema can be split into out-of plane and in-plane contributions by the symmetry
	of angular momenta matrix elements between the unoccupied and occupied $d$-states near the Fermi level. Essentially, the substrate-mediated coupling of Co adatoms to each other 
	ever so slightly alters the occupancies of $d$ orbitals of the adatoms resulting in the change of local magnetic anisotropy magnitude and even its direction.

	This being said, it should not be any more surprising that the exchange interaction between Co adatoms has a pronounced effect on their magnetic anisotropy. Indeed, within a 
	generalized Heisenberg Hamiltonian model the \Ema can be split into on-site and inter-site contributions and renormalized to an effective second-order magnetic anisotropy 
	 which contains both contributions that appear from the anisotropy of the inter-site
        exchange interactions underlining their relationship \cite{Weinberger2008,Aas2013}. From the electronic point of view, on can also look at the problem in the frame of a 
	simple Anderson model picture \cite{Oswald1985}. According to it, FM coupling results in the atomic levels of the adatoms being split in both spin channels, while AFM coupled 
	impurities exhibits single-atom-like electronic structures. This leads the above mentioned slight changes in the occupancies and therethrough to the apperance of the non-local 
	anisotropy contribution. This dependence of \Ema on the exchange order is well traceable in our calculations. The blue triangles in Figure~\ref{fig:02:maepol}(a) denote the 
	$d$-dependent \Ema values for AFM-coupled Co dimers on Cu(001). The trends of both FM and AFM curves are very close, yet the AFM curve unmistakeable exhibits less variation 
	in \Ema values in line with the above electronic-structure arguments. Generally, the presence of interatomic coupling tends to increase the \Ema values per atom as compared 
	to an isolated atom (for Co \Ema values of $0.45~\unt{meV}$ per atom in a dimer in average as compared to $0.26~\unt{meV}$ for a single adatom).

	The dependence of anisotropy of coupled Co-atoms on Pt(001) surface on the separation $d$ and their magnetic ordering is even more feature-full (Figure~\ref{fig:02:maepol}(b)). 
	Not only does the anisotropy exhibit a highly non-monotonous dependence on $d$ featuring an oscillatory switching between in-plane and out-of-plane anisotropies. It 
	also shows a much stronger dependence on the magnetic ordering of the adatom spins, a change of coupling sign usually bringing about the change of the sign of \Ema. These 
	contrasting results reveal the di\-ffe\-rent underlying adatom-substrate interaction mechanisms. Essentially, the interplay between surface-mediated and strong spin-orbit 
	interactions (existing in the later case) together with the magnetic order of the dimers determine the \Ema nature. First and foremost, the strong hybridization between 
	the Co-3$d$ with the Pt-5$d$ states results in the larger values of \Ema for Co/Pt in comparison with Co/Cu. The striking susceptibility of anisotropy of Co 
         dimers on Pt(001) as compared to Cu(001) can be understood if one considers that Pt-surface is a much more polarizable substrate of the two, as evidenced, f.e., by the 
         strong directionality of RKKY thereon \cite{Zhou2010} and long distance character of the interaction even in absence of, e.g., a surface state \cite{Stepanyuk2004,Wiebe2011}.
         The difference in dimer-induced substrate polarizations is shown in Figure~\ref{fig:02:maepol}(c) and (d) for FM and AFM coupled Co dimers at $d=6.28$~{\AA}. While on Pt(001) Co 
         atoms show show robust and large magnetic moments of $\mu_{\mathrm Co}=2.2\,\mu_B$ and a spreading substrate polarization cloud, on Cu(001) the magnetic moment of Co adatoms 
         is $\mu_{\mathrm Co}=1.9\,\mu_B$ and no induced polarization in the surface was observed. 
	
	It is thus clear that anisotropy is not, as it is often believed, an intrinsic constant of a single adsorbate and should be treated with care. While it cannot be unambiguously claimed, that interaction with surrounding magnetic adatoms is solely responsible for the dispersion of experimentally estimated anisotropy values \cite{Oberg2014}, it should definitely be duly taken into account when interpreting the results of past and future experiments.

%\section{Separation dependent response to magnetic fields}

	Another thing worth mentioning is that many experimentally observable magnetic characteristics of single adatoms and atomic assemblies \cite{Wiebe2011,Khajetoorians2012} are 
	intimately linked to such parameters as exchange interaction and anisotropy. To illustrate the implication of the dependence of magnetic anisotropy in surface structures on
	the interatomic interactions we simulate the response of Co dimers at different interatomic separations $d$ to the applied external magnetic field $\vec{B}$ as described by 
	single-atom magnetization curves \cite{Wiebe2011,Khajetoorians2012}. The simulation is carried out using the stochastic Kinetic-Monte-Carlo approach which is known to yield 
	accurate results in good agreement with experimental observations \cite{Smirnov2009,Li2006}. The basic characteristics of the system hereby are the remanence and coercivity. 
	They are also tightly correlated with relaxation times of spin systems \cite{Balashov2009,Miyamachi2013}, such as single adatoms and SMMs.

	To keep our simulations as close as possible to those used for fitting experimentally observed magnetization curves we describe our Co dimers on Cu or Pt surfaces with a 
	classical Heisenberg Hamiltonian
	\begin{equation}
		H = J(d)\;\vec{S}_1\vec{S}_2 - \sum_{i=1,2} K_i(d) \; S_{i,z}^2 - \mu \sum_{i=1,2} \vec{B} \vec{S_i},\label{eqn:ham}
	\end{equation}
	where the exchange coupling $J(d)$ and \Ema constant $K(d)$ are extracted from our \textit{ab intio} calculations. For simplicity, the anisotropy is considered to be uniaxial
	and the external magnetic field $\vec{B}$ to be always collinear to the easy anisotropy axis. The magnitude of the spins $\abs{\vec{S}_i}$ is considered constant with respect
	to interatomic separation and correspond to a magnetic moment of $m_{Co} = 1.92~\mB$ as yielded by our first-principle calculations. The temperature of the simulation is taken
	to be $0.4~\unt{K}$ which is within the typical experimental range \cite{Meier2008}.

	\begin{figure}[t]
		\centering\includegraphics{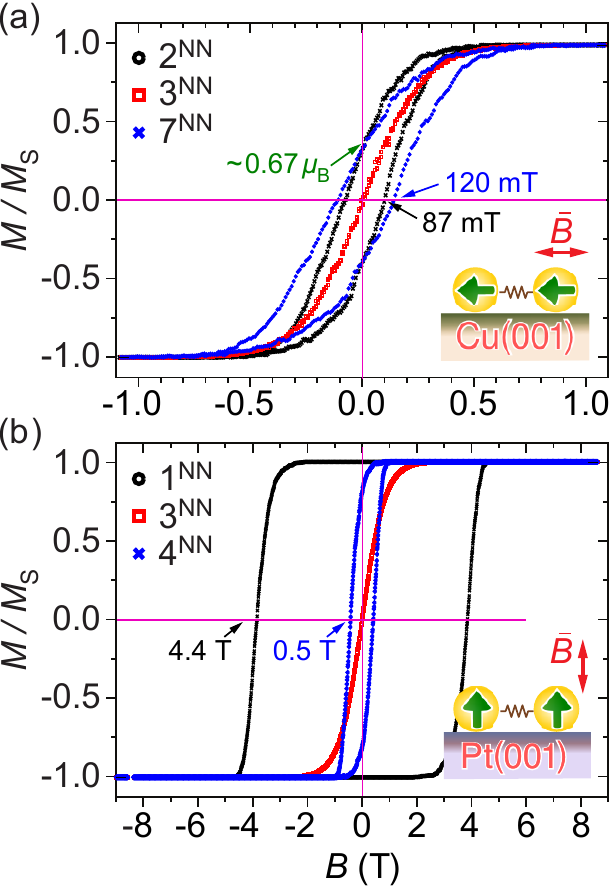}
		\caption{Hysteresis loops ($M(B)/M_\mathrm{S}$) for representative Co dimers on Cu (a) and Pt (b) surfaces. The magnetization $M$ is normalized by its saturation value $M_\mathrm{S}$.} 
		\label{fig:03:hist}
	\end{figure}

	Figure~\ref{fig:03:hist} shows magnetization curves for several representative Co dimers on Cu(001) (a) and Pt(001) (b) substrates. The magnetic field is swept from $-B_0$ to $+B_0$ and back ($B_0 = 2\unt{T}(10\unt{T})$ for Co dimers on Cu(Pt), respectively) by increments of $1~\unt{mT}$ making full cycles at a sweeping rate of $dB/dt = 130~\unt{T \times sec^{-1}}$ (according to Ref.~\citenum{Li2006}). On Cu(001) surface Co dimers exhibit hysteresis loops with relatively small coercive fields of $87$ and $120~\unt{mT}$ and remanences of about $0.67~\mB$ for $d=3.41$ and $8.11~\AA$, respectively. It is apparent, however, that the hysteresis walls are differently sloped hinting at a variation of the defining magnetic parameters. The dimer with $d = 5.17~\AA$, on the contrary, displays a paramagnetic response. This behavior is consistent with the idea of the first and second dimers having a combination of stronger exchange coupling and weaker per-atom anisotropies, or vice-versa, strong anisotropy but weak 
exchange coupling (see Figures~\ref{fig:01:exch} and \ref{fig:02:maepol}(a)), which defines the shapes of the hysteresis loops. The third dimer with negligible exchange, though relatively strong anisotropy, behaves as two independent adatoms, which are known to exhibit paramagnetic behavior due to quantum tunneling if magnetization \cite{Smirnov2009}. A similar behavior is found in the case of the Co dimers on Pt. If the exchange interaction is small ($\sim 3~\unt{meV}$), no magnetic response is found as in the case of the dimer with $d=5.63~\AA$ (Figure~\ref{fig:03:hist}(b)). Moreover, one observes that the larger \Ema values obtained for these systems prompt higher coercive fields (up to $4.4~\unt{T}$). One can again conclude then that strong exchange interactions are responsible for the hysteresis stabilization while \Ema controls the shape of the magnetization curves. The interplay of \Ema and exchange interaction defining the shape of the hysteresis loops of single atom magnetization curves can be of 
import when analyzing the the results of corresponding experimental studies \cite{Wahl2007,Pruser2014}, though in many cases the impact of the interatomic interaction shall be somewhat 
weaker, especially when the adatoms are placed on semi-insulating substrates or more sparsely distributed on a surface \cite{Meier2008,Khajetoorians2012}. \\
To draw a bottomline, with our study we underline the importance of departing from the oversimplified picture of anisotropy being a purely local constant and considering the apparent 
influence thereon of the interatomic interaction present especially in ensembles of adatoms deposited on substrates exhibiting RKKY features. This generally non-trivial separation dependence of the anisotropy can have a appreciable effect on such characteristics of magnetic systems as Kondo behavior, quantum spin tunneling and the resulting observable quantities as the inelastic tunneling spectra and the single-atom magnetization curves.

%%%%%%%%%%%%%%%%%%%%%%%%%%%%%%%%%%%%%%%%%%%%%%%%%%%%%%%%%%%%%%%%%%%%%
%% Acknowledgement
%
%\begin{acknowledgement}
{\it Acknowledgments.-} 
The authors thank Oleg Brovko for useful discussions and helpful hints as to the improvement of the manuscript. The work was supported by the Deutsche Forschungsgemeinschaft (DFG) through SFB 762 and the project ``Structure and magnetism of cluster ensembles on metal surfaces: Microscopic theory of the fundamental interactions''.
%\end{acknowledgement}

%%%%%%%%%%%%%%%%%%%%%%%%%%%%%%%%%%%%%%%%%%%%%%%%%%%%%%%%%%%%%%%%%%%%%

\bibliography{manuscript_v2}

\end{document}